\documentclass[twocolumn,prl]{revtex4}

\begin{document}

\title{
Some clarifications on cond-mat/0508763 by M. I. Katsnelson
}

\author{M. Feldbacher, R. Arita, and K. Held}

\affiliation
{Max-Planck-Institut f\"ur Festk\"orperforschung, Heisenbergstra\ss
e 1, D-70569 Stuttgart}

\author{F.\ F.\ Assaad}

\affiliation{Universit\"at W\"urzburg, Institut f\"ur Theoretische Physik I, Am
Hubland, 97074 W\"urzburg}

\begin{abstract}
Katsnelson submitted his Comment on "Projective Quantum Monte Carlo
Method for the Anderson Impurity Model and its Application to Dynamical Mean
Field Theory" to Phys. Rev. Lett. in May 2005. We proved in our report 
that this comment was incorrect since there is no orthogonality catastrophe 
for our calculation in Phys. Rev. Lett. 93, 136405 (2004)
\cite{Feldbacher}
which is for half-filling. Now in cond-mat/0508763 \cite{Katsnelson}, Katsnelson incorporates
our proof of the invalidity of his original Comment, based on Friedel's sum rule.
Instead he now claims that the projective quantum Monte Carlo method is "unpractical"
off half-filling, overlooking that our calculations off half-filling \cite{Arita}
employ in practice a noninteracting trial Hamiltonian with the same electron density
as the interacting Hamiltonian so that there is again no orthogonality catastrophe.

{\em Note added.} In the  revised version of his comment \cite{Katsnelson2},
Katsnelson gives proper  credit to our proof.
In our reply \cite{reply}, we present the 
original proof based on the Friedel sum rule.
Moreover, we show that the orthogonality catastrophe does not affect
our results. Katsnelson's objection is not valid.
\end{abstract}

\date{March 27, 2005}
\maketitle

\end{document}